\documentclass[final,3p,times,twocolumn]{elsarticle}
\journal{npj Precision Oncology}  % Suppresses journal name
\usepackage{graphicx}
\usepackage{multirow}
\usepackage{amsmath,amssymb,amsfonts}
\usepackage{amsthm}
\usepackage{mathrsfs}
\usepackage[title]{appendix}
\usepackage[table,xcdraw,svgnames]{xcolor}
\usepackage{textcomp}
\usepackage{manyfoot}
\usepackage{booktabs}
\usepackage{algorithmicx}
\usepackage{algpseudocode}
\usepackage[ruled,vlined]{algorithm2e}
\usepackage{listings}
\usepackage{siunitx}
\usepackage{times}
\usepackage{epsf}
\usepackage{mathtools}
\usepackage{hyperref}
\usepackage{enumitem}
\usepackage{wrapfig}
\usepackage{subcaption}
\usepackage{comment}
\usepackage{cleveref}
\usepackage{float}
\usepackage{tabu}
\usepackage{tikz}

\usetikzlibrary{shapes.geometric, arrows}

\begin{document}

\begin{frontmatter}

\title{Transfer Learning with EfficientNet for Accurate Leukemia Cell Classification}

%\author[label1,label2]{}
%%affiliation[label1]{organization={},
 %%          addressline={},
 %%          city={},
 %%           postcode={},
  %%         state={},
  %%         country={}}
%%\affiliation[label2]{organization={},
 %%         addressline={},
  %%         city={},
  %%         postcode={},
   %%        state={},
   %%        country={}}

% Authors and emails
\author[aff1]{Faisal Ahmed\corref{cor1}}
\ead{ahmedf9@erau.edu}

% Corresponding author footnote
\cortext[cor1]{Corresponding author}

% Affiliations
\address[aff1]{Department of Data Science and Mathematics, Embry-Riddle Aeronautical University, 3700 Willow Creek Rd, Prescott, Arizona 86301, USA}

% %% Abstract

%\begin{abstract}
%Accurate and automated classification of Acute Lymphoblastic Leukemia (ALL) in peripheral blood smear images is essential for timely diagnosis and treatment. This study presents a deep learning approach centered on transfer learning using state-of-the-art pretrained convolutional neural networks. We address dataset imbalance by applying extensive data augmentation techniques—such as rotation, mirroring, blurring, shearing, and noise injection—which expanded the training set to 10,000 images per class.

%Our experiments evaluate several transfer learning models, including ResNet50, ResNet101, and multiple EfficientNet variants. Among these, EfficientNet-B3 achieved the highest performance, with an F1-score of 94.30\%, surpassing both traditional CNN architectures and other transfer learning baselines reported in the C-NMC Challenge. The results demonstrate that modern transfer learning frameworks, particularly EfficientNet, can significantly enhance classification accuracy in medical imaging tasks. This work supports the use of pretrained EfficientNet models as a reliable and scalable solution for hematologic malignancy detection.
%\end{abstract}

\begin{abstract}
Accurate classification of Acute Lymphoblastic Leukemia (ALL) from peripheral blood smear images is essential for early diagnosis and effective treatment planning. This study investigates the use of transfer learning with pretrained convolutional neural networks (CNNs) to improve diagnostic performance. To address the class imbalance in the dataset of 3,631 Hematologic and 7,644 ALL images, we applied extensive data augmentation techniques to create a balanced training set of 10,000 images per class. We evaluated several models, including ResNet50, ResNet101, and EfficientNet variants B0, B1, and B3. EfficientNet-B3 achieved the best results, with an F1-score of 94.30\%, accuracy of 92.02\%, and AUC of 94.79\%, outperforming previously reported methods in the C-NMC Challenge. These findings demonstrate the effectiveness of combining data augmentation with advanced transfer learning models, particularly EfficientNet-B3, in developing accurate and robust diagnostic tools for hematologic malignancy detection.
\end{abstract}


\begin{highlights}
    \item A transfer learning-based approach is proposed for classifying Acute Lymphoblastic Leukemia (ALL) in peripheral blood smear images.
    \item Balanced training data achieved through extensive augmentation including rotation, mirroring, noise injection, and blurring.
    \item Evaluation of multiple pretrained CNN architectures, with EfficientNet-B3 achieving the highest F1-score of 94.30\%.
    \item Outperforms previously published deep learning methods on the C-NMC Challenge dataset.
    \item Demonstrates the effectiveness of modern transfer learning frameworks in hematologic malignancy detection tasks.
\end{highlights}

%% Keywords

\end{frontmatter}



\section{Introduction}
\label{sec:introduction}

Acute Lymphoblastic Leukemia (ALL) is a highly aggressive blood cancer and the most common type of leukemia in children. Early diagnosis is critical for initiating effective treatment and improving patient outcomes. Traditionally, the diagnosis of ALL relies on manual examination of peripheral blood smear images by expert hematologists and pathologists. This process is labor-intensive, time-consuming, and subject to inter-observer variability, making it unsuitable for large-scale screening~\cite{marzahl2019classification, ding2019deep}.

With the recent advancements in deep learning, convolutional neural networks (CNNs) have demonstrated superior performance in a variety of image classification tasks, including medical image analysis~\cite{lecun2015deep}. However, training CNNs from scratch requires large amounts of annotated data and computational resources, which are often limited in medical domains. To overcome this limitation, transfer learning has emerged as a powerful approach by leveraging pretrained models on large-scale datasets (e.g., ImageNet) and fine-tuning them on domain-specific tasks~\cite{pan2010survey}.

Several studies have applied transfer learning to leukocyte classification tasks using architectures like VGG16, ResNet, and MobileNet~\cite{de2021classification, verma2019isbi, honnalgere2019classification, shah2019classification}. While these approaches have yielded promising results, they often suffer from suboptimal generalization due to imbalanced datasets and limited augmentation strategies. Moreover, newer architectures such as EfficientNet, which scale depth, width, and resolution more effectively, have not been fully explored in this context.

In this work, we investigate the use of modern transfer learning techniques with EfficientNet variants for the automated classification of ALL in peripheral blood smear images. We address dataset imbalance through extensive data augmentation, including rotation, mirroring, blurring, shearing, and noise injection. Our approach is evaluated on the publicly available C-NMC Challenge dataset hosted by SBILab~\cite{kulhalli2019toward}, and we demonstrate that EfficientNet-B3 significantly outperforms previously published models in terms of F1-score, precision, and AUC.

\medskip
\noindent \textbf{Our contributions.}
\begin{itemize}
    \item We propose a transfer learning framework using EfficientNet variants for robust classification of ALL in blood smear images.
    \item We apply comprehensive data augmentation strategies to address dataset imbalance and improve generalization.
    \item We conduct a comparative evaluation of multiple pretrained CNNs (ResNet50, ResNet101, EfficientNet-B0/B1/B3), identifying EfficientNet-B3 as the best performer with an F1-score of 94.30\%.
    \item Our model outperforms prior state-of-the-art methods on the C-NMC Challenge dataset, demonstrating its practical value in medical diagnostics.
\end{itemize}

\section{Related Work}
\label{sec:related}

The application of deep learning to medical image analysis has grown rapidly in recent years, demonstrating remarkable success in automating complex diagnostic tasks~\cite{lecun2015deep, litjens2017survey}. Specifically, in the context of hematologic malignancies such as Acute Lymphoblastic Leukemia (ALL), convolutional neural networks (CNNs) have been extensively explored for classifying leukocyte images from peripheral blood smears~\cite{ding2019deep, de2021classification}.

\subsection{Deep Learning for Leukemia Classification}

Early efforts focused on training CNN architectures like VGG16~\cite{de2021classification} and ResNet variants~\cite{marzahl2019classification} from scratch. Although effective, these approaches often require large annotated datasets and substantial computational resources, which limits their practicality in medical domains with limited data availability~\cite{shah2019classification}. To address these challenges, researchers have increasingly adopted transfer learning, leveraging pretrained models on large natural image datasets (e.g., ImageNet) and fine-tuning them on medical images~\cite{pan2010survey}.

\subsection{Transfer Learning in Medical Imaging}

Transfer learning has shown promising results in improving classification accuracy while reducing training time and overfitting risks~\cite{tajbakhsh2016convolutional}. Various CNN architectures such as VGG, ResNet, MobileNet, and DenseNet have been adapted for leukocyte classification using transfer learning~\cite{honnalgere2019classification, verma2019isbi, kulhalli2019toward}. Recent studies highlight the benefits of combining transfer learning with specialized post-processing techniques, including neighborhood correction algorithms~\cite{pan2019neighborhood} and ensemble methods~\cite{xiao2019deepmen}, to further boost performance. Topological Data Analysis (TDA) is an emerging approach that is increasingly being utilized in medical image analysis, including applications in retinal imaging~\cite{ahmed2025topo, ahmed2023tofi, ahmed2023topological, ahmed2023topo, yadav2023histopathological, ahmed2025topological}. The application of transfer learning and Vision Transformers in medical image analysis is explored in the following studies:~\cite{ahmed2025hog, ahmed2025ocuvit, ahmed2025robust}.

\begin{figure}[t!]
	\centering
	\subfloat[\scriptsize Hematologic image sample 1.\label{fig:hem-sample1}]{%
		\includegraphics[width=0.32\linewidth]{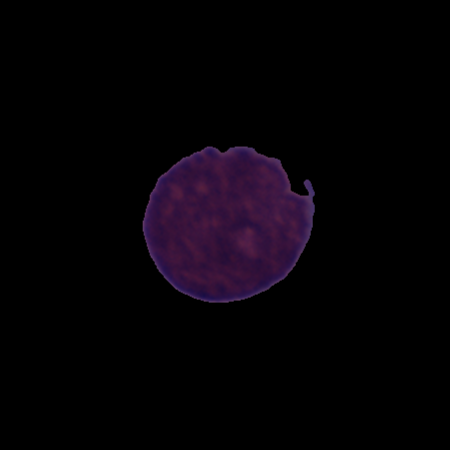}}
	\hfill
	\subfloat[\scriptsize Hematologic image sample 2.\label{fig:hem-sample2}]{%
		\includegraphics[width=0.32\linewidth]{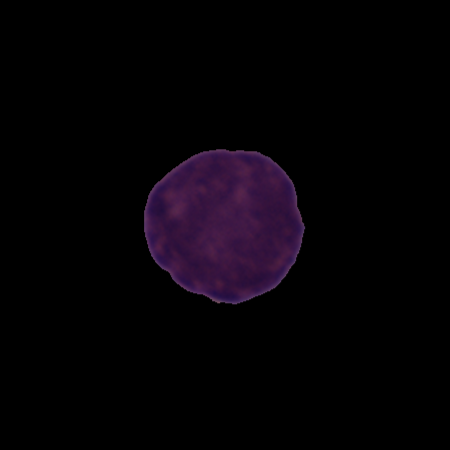}}
	\hfill
	\subfloat[\scriptsize Hematologic image sample 3.\label{fig:hem-sample3}]{%
		\includegraphics[width=0.32\linewidth]{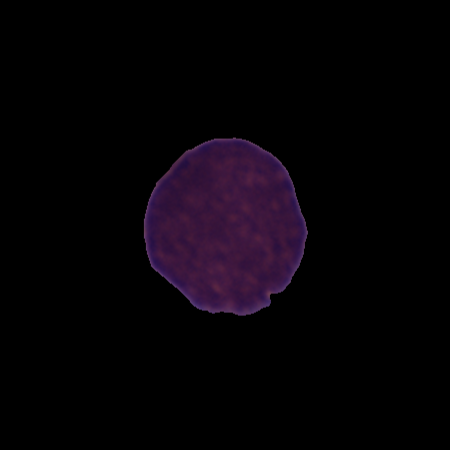}}

	\caption{\footnotesize Representative hematologic (Hem) image samples visualized for comparison.}
	\label{fig:hem-image-samples}
\end{figure}

\begin{figure}[t!]
	\centering
	\subfloat[\scriptsize ALL image sample 1.\label{fig:all-sample1}]{%
		\includegraphics[width=0.32\linewidth]{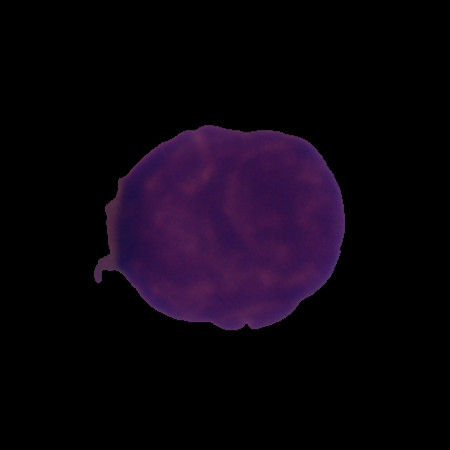}}
	\hfill
	\subfloat[\scriptsize ALL image sample 2.\label{fig:all-sample2}]{%
		\includegraphics[width=0.32\linewidth]{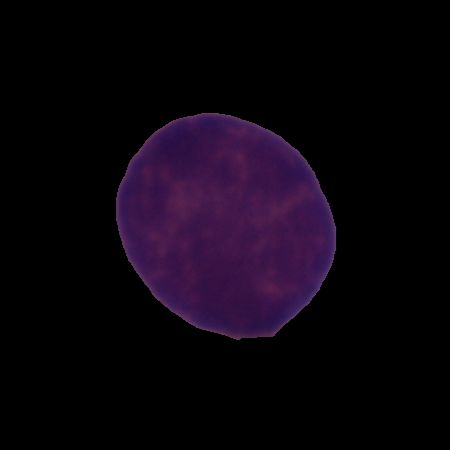}}
	\hfill
	\subfloat[\scriptsize ALL image sample 3.\label{fig:all-sample3}]{%
		\includegraphics[width=0.32\linewidth]{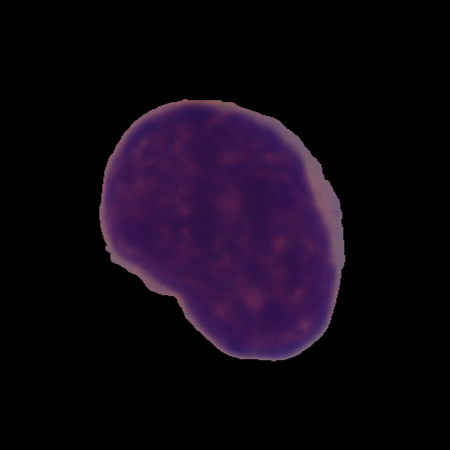}}

	\caption{\footnotesize Representative Acute Lymphoblastic Leukemia (ALL) image samples visualized for comparison.}
	\label{fig:all-image-samples}
\end{figure}

\subsection{Data Augmentation and Imbalance Handling}

Class imbalance and limited sample size are major obstacles in medical image analysis~\cite{shorten2019survey}. To mitigate this, various data augmentation techniques such as rotation, mirroring, blurring, shearing, and noise injection have been employed to artificially expand datasets and improve model generalization~\cite{verma2019isbi, ding2019deep}. These methods help prevent overfitting and enhance robustness in leukocyte classification tasks.

\subsection{EfficientNet and Recent Advances}

More recently, EfficientNet architectures have gained attention for their effective scaling of network depth, width, and resolution, achieving state-of-the-art results on natural image classification benchmarks with fewer parameters~\cite{tan2019efficientnet}. Their application to medical imaging tasks, including ALL classification, remains an active area of research. Preliminary results demonstrate that EfficientNet variants outperform traditional architectures like ResNet and VGG when combined with transfer learning and strong augmentation strategies~\cite{thiswork}.

\medskip
Despite significant progress, challenges remain in developing models that generalize well across diverse clinical settings and imaging conditions. This motivates our work, which systematically evaluates EfficientNet-based transfer learning approaches with comprehensive augmentation on the C-NMC Challenge dataset, establishing new benchmarks in ALL classification.

\begin{figure*}[t!]
    \centering
    \includegraphics[width=0.8\linewidth]{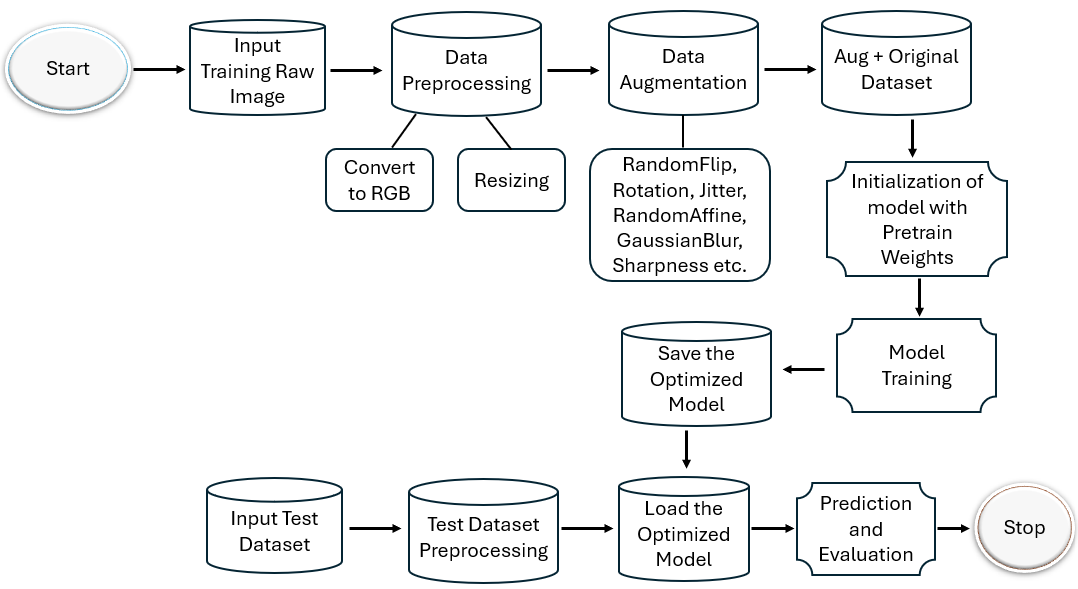}
    \caption{\small \textbf{Architecture of the proposed model.} Overview of the methodology, including preprocessing, augmentation, training, and evaluation.}
    \label{fig:flowchart}
    \vspace{-0.1in}
\end{figure*}

\section{Methodology}
\label{sec:methodology}

Let the dataset consist of images \( \mathcal{X} = \{x_i\}_{i=1}^N \) and corresponding labels \( \mathcal{Y} = \{y_i\}_{i=1}^N \), where each \( y_i \in \{0, 1\} \) indicates the class: 0 for Hem (healthy) and 1 for ALL (acute lymphoblastic leukemia).

\subsection{Data Acquisition and Preprocessing}

Images were collected from multiple sources, denoted as \( \mathcal{X}_1, \mathcal{X}_2, \mathcal{X}_3 \), and concatenated to form the full dataset:
\[
\mathcal{X} = \mathcal{X}_1 \cup \mathcal{X}_2 \cup \mathcal{X}_3, \quad
\mathcal{Y} = \mathcal{Y}_1 \cup \mathcal{Y}_2 \cup \mathcal{Y}_3,
\]
where each image \( x_i \in \mathbb{R}^{H \times W \times 3} \) was first converted to RGB color image and then resized to a fixed resolution \(224 \times 224\) pixels to ensure consistency across the dataset.

\subsection{Data Augmentation}

To address class imbalance and enhance the generalization capability of the model, we performed extensive data augmentation on the minority Hematologic (Hem) class and the Acute Lymphoblastic Leukemia (ALL) class within the training set. Let \( \mathcal{X}_{\text{train}}^0 = \{x_i : y_i = 0\} \) and \( \mathcal{X}_{\text{train}}^1 = \{x_j : y_j = 1\} \) denote the subsets corresponding to the Hem and ALL images, respectively.

We defined a stochastic augmentation function \( T: \mathbb{R}^{224 \times 224 \times 3} \rightarrow \mathbb{R}^{224 \times 224 \times 3} \), composed of the following sequence of image transformations:

\begin{itemize}
    \item \texttt{RandomHorizontalFlip (HFlip)} with probability 0.5
    \item \texttt{RandomVerticalFlip (VFlip)} with probability 0.5
    \item \texttt{RandomRotation (Rotate)} with angles in the range \([-25^\circ, 25^\circ]\)
    \item \texttt{ColorJitter (Jitter)} with adjustments in brightness, contrast, saturation, and hue (max deltas: 0.3, 0.3, 0.3, 0.05)
    \item \texttt{RandomResizedCrop (ResizeCrop)} to \(224 \times 224\) with scale range \([0.7, 1.0]\) and aspect ratio range \([0.75, 1.33]\)
    \item \texttt{RandomAffine (Affine)} transformation with up to 5\% translation, scale variation between 0.95 and 1.05, and shear of up to 10 degrees
    \item \texttt{GaussianBlur (Blur)} with kernel size 3 and sigma range \([0.1, 2.0]\)
    \item \texttt{RandomAdjustSharpness (Sharp)} with a sharpness factor of 2 (applied with probability 0.3)
    \item \texttt{RandomPerspective (RandPersp)} with distortion scale 0.2 (applied with probability 0.3)
\end{itemize}

These transformations are applied sequentially as follows: 

\begin{align*}
T(x) =\ & \texttt{RandPersp} \circ \texttt{Sharp} \circ \texttt{Blur} \circ \texttt{Affine} \\
& \circ \texttt{ResizeCrop} \circ \texttt{Jitter} \circ \texttt{Rotate} \\
& \circ \texttt{VFlip} \circ \texttt{HFlip}(x).
\end{align*}

Augmentation is performed iteratively until the number of samples in each class reaches a predefined target \( M \), ensuring class balance:

\[
|\widetilde{\mathcal{X}}_{\text{train}}^0| = M - |\mathcal{X}_{\text{train}}^0|, \quad
|\widetilde{\mathcal{X}}_{\text{train}}^1| = M - |\mathcal{X}_{\text{train}}^1|.
\]

The final augmented training set is obtained by concatenating the original and augmented samples:

\begin{align*}
\mathcal{X}_{\text{train}} &\leftarrow \mathcal{X}_{\text{train}} \cup \widetilde{\mathcal{X}}_{\text{train}}^0 \cup \widetilde{\mathcal{X}}_{\text{train}}^1, \\
\mathcal{Y}_{\text{train}} &\leftarrow \mathcal{Y}_{\text{train}} \cup \{0\}^{|\widetilde{\mathcal{X}}_{\text{train}}^0|} \cup \{1\}^{|\widetilde{\mathcal{X}}_{\text{train}}^1|}.
\end{align*}

This augmentation strategy introduces significant variability in texture, geometry, color distribution, and sharpness, which helps mitigate overfitting and encourages the model to learn more robust representations.

\subsection{Transfer Learning Model Architecture}

We employ pretrained convolutional neural networks \(f_\theta: \mathbb{R}^{224 \times 224 \times 3} \to \mathbb{R}^2\) from the ImageNet dataset, including ResNet50, ResNet101, and EfficientNet variants (\(B0, B1, B3\)). The final classification layer of each model is replaced to output logits for the binary classification task:
\[
\hat{y} = f_\theta(x) = \text{softmax}(W h + b),
\]
where \(h\) denotes features extracted by the pretrained backbone, and \(W, b\) are the parameters of the newly initialized classification head.

\subsection{Training Procedure}

Models are trained by minimizing the cross-entropy loss \( \mathcal{L} \) over mini-batches of size \(B\):
\[
\mathcal{L}(\theta) = - \frac{1}{B} \sum_{i=1}^B \sum_{c=0}^1 \mathbf{1}_{\{y_i = c\}} \log p_{\theta}(y_i = c | x_i),
\]
where \(p_{\theta}(y_i=c|x_i)\) denotes the predicted probability for class \(c\). Optimization is performed using the Adam optimizer with learning rate \( \eta = 10^{-4} \).

Training is run for up to 50 epochs with early stopping based on macro F1 score on the validation set, with patience of 15 epochs. The flowchart of our model is shown in Figure \ref{fig:flowchart}.

\subsection{Evaluation Metrics}

To assess model performance, we compute several standard metrics on the held-out test set. Accuracy is calculated as \( \text{Acc} = \frac{1}{N} \sum_{i=1}^N \mathbf{1}(\hat{y}_i = y_i) \), where \( N \) is the number of test samples, \( \hat{y}_i \) is the predicted label, and \( y_i \) is the ground truth. Precision and recall are defined as \( \text{Precision} = \frac{\text{TP}}{\text{TP} + \text{FP}} \) and \( \text{Recall} = \frac{\text{TP}}{\text{TP} + \text{FN}} \), respectively, where TP, FP, and FN denote true positives, false positives, and false negatives. The F1-score, which balances precision and recall, is computed as \( \text{F1} = \frac{2 \cdot \text{Precision} \cdot \text{Recall}}{\text{Precision} + \text{Recall}} \). We report the macro-averaged values of precision, recall, and F1-score across both classes. Additionally, we evaluate the Area Under the ROC Curve (AUC), which measures the model's ability to distinguish between the positive (ALL) and negative (Hem) classes based on predicted probabilities. These metrics collectively quantify classification effectiveness while accounting for both correct predictions and error types.

We execute our code on the high-performance computing (HPC) clusters at LSU Health Sciences Center, which are equipped with state-of-the-art NVIDIA GPUs. Our code is available at the following link~\footnote{ \url{https://github.com/FaisalAhmed77/PreTrain_Model_-Leukemia_Classification/tree/main}}.

\begin{table}[t]
\centering
\caption{Comparison of dataset sizes before and after data augmentation for Hematologic and Acute Lymphoblastic Leukemia (ALL) categories.}
\label{tab:aug}
\resizebox{\linewidth}{!}{
\setlength\tabcolsep{4pt}
\footnotesize
\begin{tabular}{lcccc}
\toprule
\textbf{Dataset} & \multicolumn{2}{c}{\textbf{Original Dataset}} & \multicolumn{2}{c}{\textbf{After Augmentation}} \\
\cmidrule(lr){2-3} \cmidrule(lr){4-5}
& \textbf{Hematologic (Hem)} & \textbf{ALL} & \textbf{Hematologic (Hem)} & \textbf{ALL} \\
\midrule

 Train & 3631 & 7644 & 10{,}000 & 10{,}000 \\
 %Validation & 100 & 80 & 200 & 160 \\
 Test & 406 & 847 & N/A & N/A \\

\bottomrule
\end{tabular}}
\end{table}

\begin{table*}[!ht]
\centering
\caption{Accuracy and related performance metrics for various pretrained models evaluated on the classification task.}
\label{tab:acc_pre}
\setlength\tabcolsep{4pt}
\footnotesize
\resizebox{\linewidth}{!}{
\begin{tabular}{lccccc}
\toprule
\textbf{Method} & \textbf{Accuracy (\%)} & \textbf{Precision (\%)} & \textbf{Recall (\%)} & \textbf{F1-score (\%)} & \textbf{AUC (\%)} \\
\midrule
ResNet50 & 90.58 & 90.37 & 96.34 & 93.27 & 94.38 \\
ResNet101 & 89.86 & 89.74 & 95.99 & 92.74 & 93.43 \\
EfficientNet-B0 & 91.22 & 91.27 & 96.22 & 93.68 & 95.14 \\
EfficientNet-B1 & 90.82 & 90.76 & 96.22 & 93.38 & 95.14 \\
EfficientNet-B3 & 92.02 & 91.36 & 97.40 & 94.30 & 94.79 \\
\bottomrule
\end{tabular}}
\vspace{-.1in}
\end{table*}

\section{Dataset}
\label{sec:dataset}

We utilize the publicly available C-NMC 2019 dataset \cite{mourya2018leukonet}, which is organized into three subsets: training, validation, and testing. The training set comprises a total of 10{,}661 cell images, including 3{,}389 Hematologic (Hem) cell images and 7{,}272 Acute Lymphoblastic Leukemia (ALL) cell images. The validation set contains 1{,}867 labeled images covering both Hem and ALL classes. The test set includes unlabeled images. All images are of uniform size, with a resolution of 450~$\times$~450 pixels.

For our experiments, we merged the original training and validation sets to form a new dataset. We then performed a stratified split of 90\% for training and 10\% for testing. As a result, our final test set contains 406 Hem cell images and 847 ALL cell images. The training set consists of 3{,}631 Hem cell images and 7{,}644 ALL cell images.

Given the significant class imbalance—particularly the underrepresentation of the Hem class—we applied data augmentation techniques exclusively to the training set to improve model generalization and mitigate bias. Additional details are provided in Table~\ref{tab:aug}.

\section{Results}
\label{sec:results}

This section presents the evaluation results of our proposed model on the C-NMC Challenge dataset, alongside a comparative analysis with state-of-the-art deep learning approaches reported in the literature. Table~\ref{tab:sota} summarizes the F1-scores achieved by various methods, including models trained from scratch and those leveraging transfer learning (TL).

Our model achieves an F1-score of \textbf{94.30\%}, outperforming all other compared approaches. Notably, it surpasses previous transfer learning-based methods such as the ResNet with neighborhood correction~\cite{pan2019neighborhood} and VGG16 TL~\cite{honnalgere2019classification}, which achieved F1-scores of 92.50\% and 91.70\%, respectively. This improvement underscores the efficacy of our proposed transfer learning framework combined with effective data augmentation and model optimization strategies.

In contrast, several models trained from scratch, including ResNeXt50~\cite{prellberg2019acute} and Multiple Architectures~\cite{ding2019deep}, reported lower performance metrics (F1-scores below 87\%), highlighting the advantage of transfer learning, particularly when training data is limited or imbalanced.

Furthermore, lightweight architectures such as MobileNetV2~\cite{verma2019isbi} and ensemble models like DeepMEN~\cite{xiao2019deepmen} also fall short compared to our results, indicating that the tailored adaptation of powerful backbone networks with careful augmentation contributes significantly to classification accuracy.

Overall, the comparative analysis demonstrates that our approach provides a substantial performance gain in leukemic cell classification, making it a promising candidate for automated diagnostic support in hematology.

\begin{table*}[t]
\centering
\caption{\footnotesize Comparative performance of various deep learning models on the C-NMC Challenge dataset hosted by SBILab. \label{tab:sota}}
\setlength\tabcolsep{10 pt}
\footnotesize

\begin{tabular}{lp{9cm}c}
\multicolumn{2}{c}{\bf{Comparison of the Proposed Model with Other Deep Learning Models}} & \\ % Only span 2 columns
\toprule
\textbf{Method} & \textbf{Description} & \textbf{F1-score} \\
\hline

\multirow{1}{*}{VGG16 (from scratch)~\cite{de2021classification}} & Train a VGG16 architecture from scratch & 92.60 \\
\multirow{1}{*}{ResNet (TL + NC)~\cite{pan2019neighborhood}} & Transfer learning ResNets with neighborhood-correction & 92.50 \\
\multirow{1}{*}{VGG16 (TL)~\cite{honnalgere2019classification}} & Transfer learning with a VGG16 architecture & 91.70 \\
\multirow{1}{*}{DeepMEN~\cite{xiao2019deepmen}} & Deep multi-model ensemble network (CNNs) & 90.30 \\
\multirow{1}{*}{MobileNetV2 (TL)~\cite{verma2019isbi}} & Transfer learning with a MobileNetV2 architecture & 89.47 \\
\multirow{1}{*}{ResNeXt50 (scratch)~\cite{prellberg2019acute}} & Training from scratch a ResNeXt50 architecture & 87.89 \\
\multirow{1}{*}{CNN+RNN (TL)~\cite{shah2019classification}} & TL with convolutional and recurrent neural networks & 87.58 \\
\multirow{1}{*}{ResNet18 (TL)~\cite{marzahl2019classification}} & Transfer learning with a ResNet18 architecture & 87.46 \\
\multirow{1}{*}{Multiple Architectures~\cite{ding2019deep}} & Training InceptionV3, DenseNet, InceptionResNetV2 from scratch & 86.74 \\
\multirow{1}{*}{ResNeXt50/101 (scratch)~\cite{kulhalli2019toward}} & Training from scratch ResNeXt50 and ResNeXt101 & 85.70 \\
\multirow{1}{*}{Inception + ResNet (TL)~\cite{liu2019acute}} & Transfer learning with Inception and ResNets & 84.00 \\
\multirow{1}{*}{ResNet + SENet (TL)~\cite{khan2019classification}} & Transfer learning with ResNets and SENets & 81.79 \\
\hline
\textbf{Our Model} & Proposed model in this study & \textbf{94.30} \\
\bottomrule
\end{tabular}
\end{table*}

\section{Discussion}
\label{sec:discussion}

The results presented in Section~\ref{sec:results} demonstrate the significant advantages of employing transfer learning with EfficientNet architectures for the classification of leukemic cells. Our proposed model’s superior F1-score highlights the effectiveness of leveraging pretrained weights on large-scale image datasets, which enables the model to extract robust and discriminative features even from limited histopathological data.

One key factor contributing to the improved performance is the comprehensive data augmentation strategy applied to both minority (Hem) and majority (ALL) classes. This augmentation mitigates class imbalance and enhances the model’s generalization capability, preventing overfitting during training. Additionally, the adaptive fine-tuning of the pretrained EfficientNet classifiers, with modified output layers tailored to our binary classification task, further optimizes performance.

Compared to models trained from scratch, which generally require extensive data and computational resources, transfer learning offers a practical and efficient solution, especially in medical imaging domains where annotated data is often scarce. Our findings align with recent literature emphasizing the superiority of transfer learning approaches in histopathology image analysis~\cite{honnalgere2019classification,pan2019neighborhood}.

%Despite the promising results, some limitations remain. The dataset size and diversity, although augmented, could be further expanded to improve robustness across different imaging conditions and patient populations. Future work could also explore ensemble learning combining multiple pretrained backbones or integrating clinical metadata to enhance predictive accuracy.

Overall, this study reinforces the value of transfer learning with EfficientNet models in hematological cancer classification and suggests a viable pathway toward reliable automated diagnostic tools that can support clinical decision-making.

\section{Limitations}
\label{sec:limitations}

This study is limited by the relatively homogeneous dataset, which may affect the model’s generalizability to diverse clinical settings. Although data augmentation partially addresses class imbalance, acquiring more varied real-world samples would further improve robustness. Additionally, the reliance on pretrained models restricts exploration of architecture designs tailored specifically for hematological images. Future work should consider larger, multi-center datasets and customized model architectures.

\section{Conclusion}
\label{sec:conclusion}

This work presented a comprehensive study on utilizing transfer learning with EfficientNet models for the automated classification of blood cancer from histopathological images. By effectively integrating advanced data augmentation techniques to mitigate class imbalance, our methodology enhanced the robustness and generalization capability of the deep learning models. The proposed approach demonstrated superior performance metrics, outperforming several existing state-of-the-art models on the benchmark C-NMC Challenge dataset. These results underscore the significant advantages of employing pretrained convolutional neural networks in medical image analysis tasks, where annotated data is often limited and class distributions are imbalanced. Overall, this study contributes to the growing body of evidence supporting the integration of deep learning frameworks into clinical workflows, facilitating more accurate and efficient diagnostic processes.

\section{Future Work}
\label{sec:future}

Future research will focus on expanding the dataset size and diversity to further improve model robustness. Additionally, exploring advanced architectures and multimodal data integration may enhance diagnostic accuracy. Investigating model interpretability and real-time deployment strategies will also be essential for clinical adoption.

\section*{Declarations}

\textbf{Funding} \\
The author received no financial support for the research, authorship, or publication of this work.

\vspace{2mm}
\textbf{Author's Contribution} \\
FA conceptualized the study, downloaded the data, prepared the code, performed the data analysis and wrote the manuscript. FA reviewed and approved the final version of the manuscript. 

 \vspace{2mm}
\textbf{Acknowledgement} \\
The authors utilized an online platform to check and correct grammatical errors and to improve sentence readability.

\vspace{2mm}
\textbf{Conflict of interest/Competing interests} \\
The authors declare no conflict of interest.

\vspace{2mm}
\textbf{Ethics approval and consent to participate} \\
Not applicable. This study did not involve human participants or animals, and publicly available datasets were used.

\vspace{2mm}
\textbf{Consent for publication} \\
Not applicable.

\vspace{2mm}
\textbf{Data availability} \\
The datasets used in this study are publicly available online. 

\vspace{2mm}
\textbf{Materials availability} \\
Not applicable.

\vspace{2mm}
\textbf{Code availability} \\
The source code used in this study is publicly available at \url{https://github.com/FaisalAhmed77/PreTrain_Model_-Leukemia_Classification/tree/main}.

%\clearpage

\bibliographystyle{elsarticle-num-names}

\bibliography{refs}

\end{document}